\def\bey{\begin{eqnarray}}
\def\eey{\end{eqnarray}}
\def\be{\begin{equation}}
\def\ee{\end{equation}}
\def\ba{\begin{array}}
\def\ea{\end{array}}
\def\gm{\gamma}

\def\ld{\lambda}
\def\Ld{\Lambda}

\def\sg{\sigma}

\def\om{\omega}
\def\r{\rho}

\def\dt{\delta}

\def\pp{\partial}
\def\pp{\partial}

\bigskip
\documentclass[onecolumn,showpacs,preprintnumbers,amsmath,amssymb]{revtex4}
\usepackage{graphicx}
\usepackage{fancyhdr}
\usepackage{dcolumn}
\usepackage{bm}
\begin{document}
\preprint{ }
\title{ Mean free paths and in-medium scattering cross sections of
energetic nucleons in neutron-rich nucleonic matter within the
relativistic impulse approximation}
\bigskip
\author{Wei-Zhou Jiang$^{1,2}$, Bao-An Li$^{1}$, and Lie-Wen Chen$^{1,3}$}
\affiliation{  $^1$ Department of Physics, Texas A\&M
University-Commerce, Commerce, TX 75429, USA\\ $^2$ Institute of
Applied Physics, Chinese Academy of Sciences, Shanghai 201800,
China\\$^3$ Institute of Theoretical Physics, Shanghai Jiao Tong
University, Shanghai 200240, China}
\bigskip

\begin{abstract}
\baselineskip18pt The mean free paths and in-medium scattering
cross sections of energetic nucleons in neutron-rich nucleonic
matter are investigated using the nucleon optical potential
obtained within the relativistic impulse approximation with the
empirical nucleon-nucleon scattering amplitudes and the nuclear
densities obtained in the relativistic mean field model. It is
found that the isospin-splitting of nucleon mean free paths,
sensitive to the imaginary part of the symmetry potential, changes
its sign at certain high kinetic energy. The in-medium
nucleon-nucleon cross sections are analytically and numerically
demonstrated to be essentially independent of the isospin
asymmetry of the medium  and increase linearly with density in the
high energy region where the relativistic impulse approximation is
applicable.
\end{abstract}
\maketitle \baselineskip 20.6pt

\section{Introduction}
The in-medium nucleon-nucleon (NN) cross sections and the
associated nucleon mean free paths (MFP) in neutron-rich nucleonic
matter are important for understanding both the structure of rare
isotopes and the reaction dynamics involving these nuclei. For
instance, much information about the size of neutron skins and
nucleon density profiles of neutron-rich nuclei can be obtained
from Glauber model analyses of the total reaction cross
sections\cite{Tan01}. A key input in these analyses is the isospin
dependent in-medium NN cross sections. Usually, however, in most
Glauber calculations only the isospin averaged free-space
nucleon-nucleon cross section is used. The knowledge about the
in-medium NN cross sections in neutron-rich matter will thus
affect significantly the extracted structure information about
rare isotopes. It is also critical for understanding nuclear
reactions as the in-medium NN cross sections determine the
stopping power and generally transport properties of nuclear
matter. The in-medium NN cross sections, as basic inputs of
transport models, are important in determining essentially all
dynamical observables, such as, the collective flows, isospin
diffusion, energy and momentum transfers, and particle
production~\cite{ch05,li05,liba05,ga05,pr05,zh07}. While the
in-medium NN cross sections and the nuclear equation of state
(EOS) are in principle determined by the same nuclear effective
interaction, although they are sensitive to different parts and
ranges of the interaction in the medium at different densities,
much more attention has been putting on investigating the EOS than
on the in-medium NN cross sections in neutron-rich matter.
However, it was demonstrated recently that the major theoretical
uncertainty in extracting the density dependence of the symmetry
energy of neutron-rich matter from heavy-ion reactions is our poor
knowledge about the isospin-dependent in-medium NN cross
sections\cite{li05}. It is encouraging to notice that more work is
being done recently in calculating the in-medium NN cross sections
using various approaches, such as, the relativistic and
non-relativistic Brueckner
theory~\cite{ga05,li93,li94,fu01,sa06,ja92,am94,Sch97,Gia96,ko98}
and the Green function approach~\cite{mao94,di99,li00}.
Nevertheless, most of these calculations are for in-medium NN
cross sections in symmetric nuclear matter. Very little
information is available for the isospin dependence of the NN
cross section in neutron-rich matter. While the available
calculations are indeed very useful, necessarily are the NN cross
sections and the EOS consistently calculated using the same
nuclear effective interactions.

The nucleon-nucleus optical potential plays a central role in
determining the in-medium NN cross sections and the nucleon MFP. In
particular, the imaginary part of the optical potential allows one
to extract easily the MFP and thus also the in-medium NN cross
sections. There are several possible ways to derive the optical
potential~\cite{ho94}. It can be calculated from various models
either non-relativistically or relativistically, with the
microscopic Brueckner approach, see, e.g.,
Ref.\cite{fu01,ja92,am94,ha86,ro06}. Theoretically, approximations
have to be made to render the calculations feasible. For instance,
the evaluation of the imaginary part of the optical potential
depends on the treatment of the widths of intermediate states that
are rather unclear or the nucleon polarization in the medium that is
in principle coupled to the vacuum. On the other hand, one can start
from a physically reasonable approximation for the optical potential
and determine its parameters using the experimental
data~\cite{kh06}. The relativistic impulse approximation (RIA)
actually follows this way. In the RIA, the optical potential is
obtained in a form similar to the non-relativistic $t\r$
approximation. Within the RIA, the basic ingredients of the optical
potential are the free Lorentz invariant NN scattering amplitudes
and the nuclear scalar and vector densities in nuclear
matter~\cite{mc83,sh83,cl83,mi83,mcn83}. An attractive feature of
the RIA is that the relativistic optical potentials are
experimentally constrained by the free-space NN scattering data. The
nuclear densities are calculated from the relativistic mean-field
(RMF) models that provide also a dynamical description for the
spin-orbit coupling~\cite{se86}. Along with the success of the RMF
models in describing the nuclear structure, the RIA was justified by
nicely reproducing the proton-nucleus elastic scattering data at
high energies. Of course, the limitation of the RIA is that it is
valid only at reasonably high energies in not so dense matter. As
most of the existing microscopic calculations were devoted to the
low and intermediate energy regions, the in-medium NN cross sections
at high energies remain largely unknown. Our study here of the
nucleon MFP and in-medium NN cross sections at high energies within
the RIA is thus complementary to the existing work. Moreover, given
the exciting scientific opportunities provided by the constructions
of several high energy radioactive beam facilities around the world,
the knowledge on the mean free paths and in-medium scattering cross
sections of energetic nucleons in neutron-rich nucleonic matter is
certainly necessary and useful.

The paper is organized as follows. In section~\ref{ria}, the main
formulas of the  RMF, RIA, and the calculations of the nucleon MFP's
and in-medium NN cross sections are briefly reviewed. The results and
discussions are presented in section~\ref{results}. Finally, a
summary is given in Section~\ref{summary}.

\section{A brief summary of the formalism}
\label{ria}

In the present work, we use the relativistic model
Lagrangian~\cite{ji07}
\begin{eqnarray} \label{eq:lag1}
{\cal L}&=& {\overline\psi}[i\gamma_{\mu}\partial^{\mu}-M
+g^*_{\sigma}\sigma-g^*_{\omega }
\gamma_{\mu}\omega^{\mu}-g^*_{\rho}\gamma_\mu \tau_3 b_0^\mu-e\frac{1}{2}%
(1+\tau_3)\gamma_\mu A^\mu]\psi +\frac{1}{2}(\partial_{\mu}\sigma\partial^{%
\mu}\sigma-m_{\sigma}^{*2}\sigma^{2})  \nonumber \\
&& - \frac{1}{4}F_{\mu\nu}F^{\mu\nu}+ \frac{1}{2}m_{\omega}^{*2}\omega_{\mu}%
\omega^{\mu} - \frac{1}{4}B_{\mu\nu} B^{\mu\nu}+
\frac{1}{2}m_{\rho}^{*2} b_{0\mu} b_0^{\mu}- \frac{1}{4}A_{\mu\nu}
A^{\mu\nu},
\end{eqnarray}
where $\psi,\sigma,\omega$, and $b_0$ are the fields of the nucleon, scalar,
vector, and isovector-vector mesons, with their masses $M, m^*_\sg,m^*_\om$,
and $m^*_\r$, respectively. The meson coupling constants and masses with
asterisks denote the density dependence, and they are given by the BR scaling~\cite%
{br05,song01,ji07}. The Dirac spinor in momentum space is written as
\begin{equation}  \label{eqn1}
\psi(k)=\sqrt{\frac{E^*+M^*}{2E^*}}\left(%
\begin{array}{c}
  1 \\
  \frac{{\bf \sg}\cdot {\bf k}}{E^*+M^*} \\
\end{array}%
\right)\chi_s,
\end{equation}
with $M^*=M-g^*_\sg\sg$, $E^*=\sqrt{M^*+{\bf k}^2}$, and $\chi_s$ the
Pauli spinor. The (0th-component) vector and scalar densities are
respectively give by
\begin{eqnarray}
  \r_{B,i} &=&\int^{k_{F,i}}_0\frac{d^3\!k}{(2\pi)^3}\psi^\dagger(k)\psi(k)=
  \frac{1}{3\pi^2}k_{F,i}^3  \\
  \r_{s,i} &=&\int^{k_{F,i}}_0\frac{d^3\!k}{(2\pi)^3}\psi^\dagger(k)\gm_0
  \psi(k)=  \int^{k_{F,i}}_0\frac{d^3\!k}{(2\pi)^3}\frac{M^*}{E^*},
\end{eqnarray}
with $i=p,n$. With   Dirac spinors, the NN scattering amplitudes can
be defined~\cite{mcn83}. The Lorentz invariant NN scattering
amplitude can be decomposed into five components
\begin{equation}\label{amp}
{ \cal F}=F_S+F_V\gm_1^\mu\gm_{2\mu}+F_T\sg_1^{\mu\nu}\sg_{2\mu\nu}+
 F_P\gm_1^5\gm_2^5 +F_A\gm_1^5\gm_1^\mu\gm_2^5\gm_2^\mu,
\end{equation}
where S,V,T,P and A denote the scalar, vector, tensor, pseudoscalar
and axial vector, respectively. The five components are dependent on
the squared momentum transfer ${\bf q}^2$. The scattering amplitude
in the RIA is obtained from the expectation value of a sum of single
two-body NN scatterings.   For the spin-saturated nucleus, only
survived are the scalar ($F_S$), vector ($F_V$) and tensor ($F_T$)
terms, while the scalar and vector parts dominate the scattering
amplitude. In the RIA, the optical potential in momentum space is
thus the sum of the products of the density and amplitude of
respective scalar and vector components~\cite{mc83,sh83}
\begin{equation}\label{optq}
\tilde{U}_{opt}({\bf q})=-\frac{4\pi i p_{lab}}{M} [F_S({\bf
q})\tilde{\r}_S({\bf q}) +\gm_0 F_V({\bf q})\tilde{\r}_B({\bf q})],
\end{equation}
where $p_{lab}$ and M are the laboratory momentum and mass of the
incident nucleon, respectively.  This expression is similar to the
$t\r$ approximation in the non-relativistic approximation. The
momentum dependence of the optical potential can give details of the
angular distribution of nucleon scatterings, while the total cross
section depends only on the forward scattering where ${\bf q}=0$.
Following Ref.~\cite{che05}, here we are interested in the optical
potential in nuclear matter where the densities in coordinate space
are constant. This means that a delta function $\dt({\bf q})$ is
induced so that  ${\bf q}=0$ in the scattering amplitudes in
Eq.(\ref{amp}). Therefore the optical potential in nuclear matter in
coordinate space is given as:
\begin{equation}\label{optr}
 U_{opt}=-\frac{4\pi i p_{lab}}{M} [F_S\r_S  +\gm_0 F_V \r_B],
\end{equation}
where $\r_S$ and $\r_B$ are the spatial scalar and vector densities
of nucleons, respectively. For the application to the scattering on
finite nuclei, one can resort to the local density approximation.
Similar to Ref.~\cite{che05}, here we adopt the $F_S$ and $F_V$
determined directly from the experimental NN phase
shifts~\cite{mcn83}. The optical potential is decomposed into the
following form
\begin{equation}\label{opt3}
U_{opt}=U_{S}^{tot}+\gm_0 U_0^{tot}.
\end{equation}
with $U_{S}^{tot}$ and $U_{0}^{tot}$ denoting, respectively, the
scalar and (0th-component) vector parts of the optical potential.
Since $U_{S}^{tot}$ and $U_{0}^{tot}$ are generally complex, they can
be further expressed as
\begin{equation}\label{opt4}
U_{S}^{tot}=U_{S}+iW_S,\hbox{ }U_{0}^{tot}=U_{0}+iW_0.
\end{equation}

An application of relativistic optical potentials is to evaluate the
nucleon MFP based on the dispersion relation which in the
relativistic frame is written as
\begin{equation}\label{disp}
(E_k-U^{tot}_0)^2={\bf k}^2+({M+U_S^{tot}})^2.
\end{equation}
Equivalently, it can be written in a form similar to the
non-relativistic approximation~\cite{ja81}
\begin{equation}\label{disp1}
\frac{k_\infty^2}{2M}=\frac{k^2}{2M}+U_{sep}^{tot}(E_{kin}),
\end{equation}
where the Schr\"odinger equivalent potential (SEP) is given by
\begin{equation}\label{sep}
U^{tot}_{sep}=U_S^{tot}+U_0^{tot}+\frac{1}{2M}({U_S^{tot}}^2-
{U_0^{tot}}^2) +\frac{U_0^{tot}}{M}E_{kin},
\end{equation}
with $k^2_\infty=E^2_{kin}+2ME_{kin}$, $E_k=E_{kin}+M$, and
$U^{tot}_{sep}=U_{sep}+iW_{sep}$. The real part of the symmetry
potential is
\begin{equation}\label{laner}
    U_{sym}=\frac{  U_{sep}^n- U_{sep}^p}{2\dt},
\end{equation}
with the isospin asymmetry $\dt=(\r_n-\r_p)/\r_B$. The above
$U_{sym}$ taken at the normal density is the well known Lane
potential extracted from nucleon-nucleus scatterings~\cite{la62}.
Similarly, we can define an imaginary symmetry potential as
\begin{equation}\label{lanei}
    W_{sym}=\frac{  W_{sep}^n- W_{sep}^p}{2\dt}.
\end{equation}
Using the dispersion relation (\ref{disp1}) and assuming a complex
momentum $k=k_R+ik_I$, the nucleon MFP $\ld$ is derived exactly as
~\cite{ligq93}
\begin{equation}\label{mfp0}
\ld=\frac{1}{2k_I}=\frac{1}{2}[-M(E_{kin}+\frac{E_{kin}^2}{2M}
-U_{sep}) +M((E_{kin}+\frac{E_{kin}^2}{2M}
-U_{sep})^2+W_{sep}^2)^{1/2}]^{-1/2}.
\end{equation}
A much simpler expression for the nucleon MFP can be derived by
expanding the momentum in the vicinity of $k_R$~\cite{ne81}. The
real and imaginary parts of the momentum is thus obtained as
\begin{equation}\label{momentum}
 k_R=  (E_{kin}^2+2ME_{kin}-2MU_{sep})^{1/2},\hbox{ }   k_I=-W_{sep}(
  \frac{k_R}{M}+\frac{\pp U_{sep}}{\pp k_R})^{-1}.
\end{equation}
Since there is no explicit momentum dependence in the optical
potentials with the RIA, the nucleon MFP is written as
\begin{equation}\label{mfp1}
    \ld_i=\frac{1}{2k^i_I}=-\frac{k^i_R}{2MW^i_{sep}}, \hbox{ }i=p,n.
\end{equation}
Numerically, we noted that almost the same nucleon MFP can be
obtained from either Eq.(\ref{mfp0}) or (\ref{mfp1}), while the
latter is more convenient for making analytical analyses. The
nucleon MFP can also be measured as the length of the unit volume
defined by the matter density and the NN cross section, and it is
thus expressed by~\cite{pa92}
\begin{equation}\label{mfp2}
    \ld_i=(\r_{p}\sg^*_{ip}+\r_n\sg^*_{in})^{-1},
\end{equation}
where $\r_p$ and $\r_n$ are respectively the proton and neutron
densities. The in-medium NN cross sections can be obtained by
inverting the above equation. To write the results compactly, we may
define the following two quantities,
\begin{equation}\label{mfp3}
    \tilde{\Ld}^{-1}=\frac{1}{2}(\frac{1}{\ld_n}+\frac{1}{\ld_p}), \hbox{ }
    \tilde{\ld}^{-1}=\frac{1}{2\dt}(\frac{1}{\ld_n}-\frac{1}{\ld_p}).
\end{equation}
They can be further written in terms of the imaginary parts of the
symmetry potential and the isoscalar SEP using  Eqs.(\ref{lanei}) and
(\ref{mfp1})
\begin{equation}\label{mfp4}
    \tilde{\Ld}^{-1}=\frac{2M}{k_R}\bar{W}_{sep}, \hbox{ }
    \tilde{\ld}^{-1}=\frac{2M}{k_R} W_{sym},
\end{equation}
where the $k_R^{n,p}$ are approximated by the $k_R$. This is a very
good approximation because at high energies the $U_{sym}$ is
negligible compared to the kinetic energy and the isoscalar SEP's
given by
\begin{equation}\label{uwsep}
\bar{U}_{sep}=({U}^n_{sep}+{U}_{sep}^p)/2,~~
\bar{W}_{sep}=({W}^n_{sep}+{W}_{sep}^p)/2.
\end{equation}
This point will be shown numerically in the next section. As it was
pointed out in Ref.~\cite{che05}, the $U_{sym}$ itself is isospin
independent because the difference between the neutron and proton
potentials is largely linear in isospin asymmetry. We find that such
an isospin independence retains also for the $W_{sym}$. It is
therefore understandable that the $\tilde{\Ld}$ and $\tilde{\ld}$
are essentially independent of the isospin asymmetry of the medium
at high energies where the RIA is valid. Consequently, at these high
energies the in-medium NN cross sections are also independent of the
isospin asymmetry of the medium. Of course, there is still a
difference between the neutron-proton and proton-proton
(neutron-neutron) cross sections. In terms of these isospin
independent quantities, the in-medium NN cross sections are obtained
as
\begin{equation}\label{imcs}
    \sg_{nn}^*=(\tilde{\Ld}^{-1}+\tilde{\ld}^{-1})/\r_B, \hbox{ }
\sg_{np}^*=(\tilde{\Ld}^{-1}-\tilde{\ld}^{-1})/\r_B.
\end{equation}
Here we assume that $\sigma_{nn}^*=\sigma_{pp}^*$, thus neglecting
the small charge symmetry breaking effect~\cite{li98} and the
isospin-dependent Pauli blocking effects in asymmetric nuclear
medium. The neglect of the former is also consistent with the model
Lagrangian we used in this work. We do not consider the tiny
proton-neutron mass splitting in free space. Moreover, our model
Lagrangian similar to those used in most other RMF models does not
include the scalar-isovector $\delta$ meson responsible for the
in-medium proton-neutron effective mass splitting. The same nn and pp
scattering amplitudes are thus expected to result in the equal
in-medium nn and pp cross sections. We stress here that the above
results are only applicable to the high energy region where the RIA
is valid and Pauli blocking effects are negligible. At low energies,
the $U_{sym}$ and isospin-dependent Pauli blocking effects are not
negligible, and the resulting NN cross sections will depend on the
isospin asymmetry of the medium.

\section{Results and discussions}
\label{results}

The RMF model used here is the SL1 we constructed in our previous
work~\cite{ji07}. The scalar densities used in the RIA are calculated
with the SL1. The scalar densities calculated at low densities
($\r\leq1.5\r_0$) are quite similar with various models ranging from
the best-fit models to the SL1. We refer the reader to our earlier
work~\cite{ji07} for more details about the RMF with the SL1
parameter set.

\begin{figure}[tbh]
\begin{center}
\vspace*{-10mm}
\includegraphics[width=10cm,height=10cm]{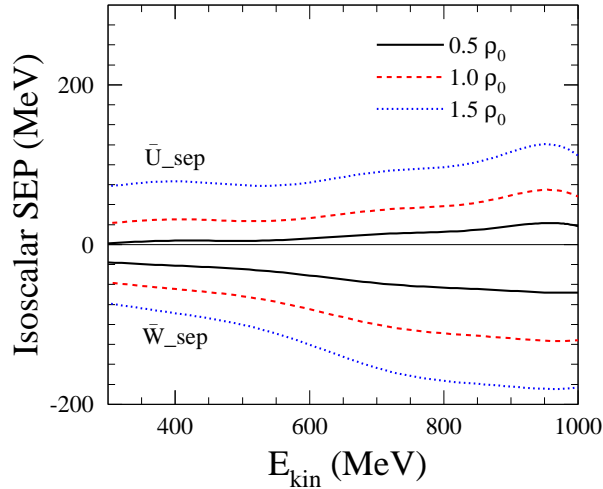}
\end{center}
\caption{(Color online) The isoscalar Schr\"odinger equivalent potential as a
function of the nucleon kinetic energy for different
densities.}\label{f:f0}
\end{figure}

\begin{figure}[tbh]
\begin{center}
\vspace*{-10mm}
\includegraphics[width=10cm,height=10cm]{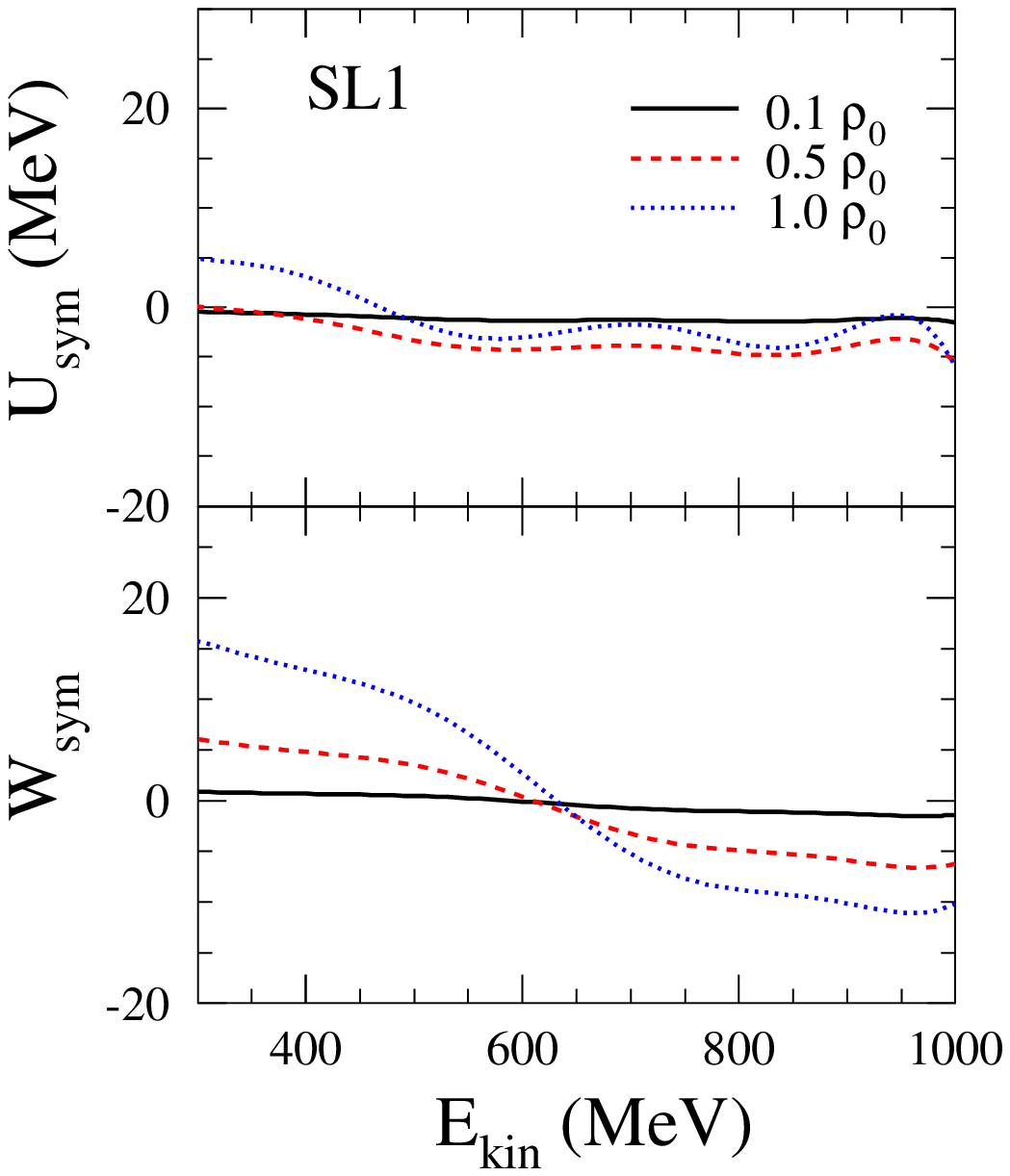}
\end{center}
\caption{(Color online) The symmetry potential as a function of the nucleon kinetic
energy for different densities.}\label{f:f1}
\end{figure}
\subsection{Schr\"odinger equivalent potential}
The proton and neutron SEP's obtained with the RIA serve as the basis
to study the nucleon MFP and in-medium NN cross sections in this
work. The isoscalar nucleon SEP, defined by the mean value of the
neutron and proton SEP's in Eq. (\ref{uwsep}), is displayed in
Fig.~\ref{f:f0}. It is seen that the imaginary SEP is stronger than
the real one for nucleons at high kinetic energies. As the kinetic
energy increases, the magnitude of the SEP increases very slowly.
Consequently, this results in a correspondingly small decrease of the
nucleon MFP, as given by Eq.(\ref{mfp1}). Shown in Fig.~\ref{f:f1}
are the real and imaginary parts of the symmetry potential, defined
by Eqs.(\ref{laner}) and (\ref{lanei}), respectively. Indeed, we find
that these potentials are independent of the isospin asymmetry of the
medium. For the $U_{sym}$, this was also found
previously~\cite{che05}. In Ref.~\cite{che05}, it was pointed out
that the real part of the symmetry potential at fixed baryon
densities is essentially a constant at high nucleon kinetic energies.
As shown in the upper panel of Fig.~\ref{f:f1}, this is also true
here. The $U_{sym}$ is negligibly small, actually very close to zero,
in the whole energy range considered. This finding is consistent with
the tendency of the energy dependence of the Lane potential extracted
from nucleon-nucleus scatterings, which indicates that the Lane
potential decreases with the increasing incident energy up to about
100 MeV above which there is still no data\cite{ho94}. The smallness
of $U_{sym}$ shown in the figure numerically justifies the
approximation of using the $k_R$ instead of the $k_R^{n,p}$ in
Eq.(\ref{mfp4}). The imaginary part of the symmetry potential, which
differentiates proton and neutron absorptions in nuclear medium,
displays a different dependence on the kinetic energy from that for
the $U_{sym}$. It is particularly interesting to note that its sign
changes around $E_{kin}\approx630MeV$ from positive to negative. This
change can be reflected in the energy dependence of the nucleon
MFP's.

\subsection{Nucleon mean free path}
\begin{figure}[tbh]
\begin{center}
\vspace*{-10mm}\includegraphics[width=10cm,height=10cm]{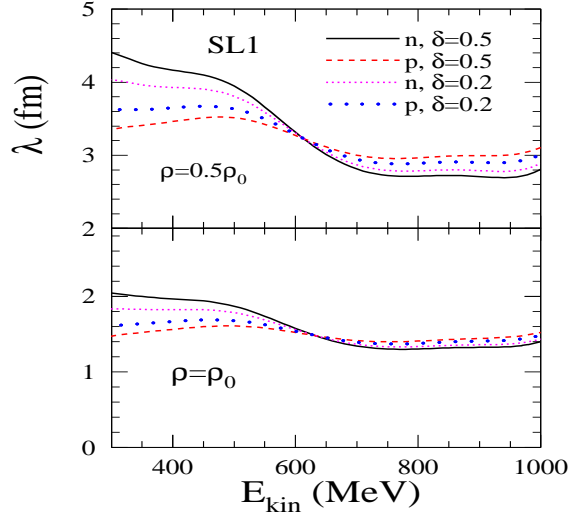}
\end{center}
\caption{(Color online) The nucleon mean free path as a function of the nucleon
kinetic energy for different densities.}\label{f:f2}
\end{figure}

The nucleon MFP's, calculated from Eq.(\ref{mfp0}) or (\ref{mfp1}),
are shown in Fig.~\ref{f:f2}. We find that at higher kinetic energies
the nucleon MFP changes little with the energy and it is much less
sensitive to the isospin asymmetry. The rapid change of the nucleon
MFP occurs around $E_{kin}\approx630$MeV. It is interesting to see
that the isospin-splitting between the neutron and proton MFP changes
at $E_{kin}\approx630$MeV. The nucleon MFP is determined by both the
real and imaginary parts of the SEP as seen in Eq.(\ref{mfp1}). The
difference between the real parts of the neutron and proton SEP's is
tiny compared to the nucleon momentum at high kinetic energies. In
fact, we see from Fig.~\ref{f:f1} that the real part, i.e.,
$U_{sym}$, is close to zero. Therefore, the change of the
isospin-splitting between the neutron and proton MFP's can be
attributed to the sign change of the imaginary part of the symmetry
potential as shown in the lower panel of Fig.~\ref{f:f1}. This change
for MFP's of neutrons and protons around 600 MeV/A may have
interesting experimental consequences and certainly deserves further
studies.

At $E_{kin}\geq700$ MeV,  the nucleon MFP is insensitive to the
isospin asymmetry of the medium. This insensitivity is due to the
fact that at high nucleon kinetic energies the $W_{sym}$, responsible
for the difference between the proton and neutron MFP's, is small
compared to the $\bar{W}_{sep}$ that dominates the contributions to
the nucleon MFP's. At lower kinetic energies, however, appreciable
sensitivity to the isospin asymmetry exists just because the
imaginary part of the symmetry potential is comparably not so small.
Comparing the two panels of Fig. 3, it is seen that the sensitivity
of the nucleon MFP to the isospin asymmetry is reduced with the
increasing density. This is also understandable because of the
decreasing nucleon MFP's and the drop of the $W_{sym}$ to
$\bar{W}_{sep}$ ratio at higher densities.

\subsection{In-medium NN cross sections}
As discussed briefly in the text below Eq.(\ref{mfp4}), in the energy
range where the RIA is valid, the in-medium NN cross sections
$\sg_{np}^*$ and $\sg_{nn,pp}^*$ are independent of the isospin
asymmetry of the medium. They depend only on the density and the
nucleon kinetic energy. Before presenting our results, it is
worthwhile to add some more discussions. Usually, the in-medium NN
cross section relies on the isospin asymmetry at low kinetic
energies. Such dependence is introduced by the isospin dependent
nucleon SEP through the real part of the momentum $k_R$, given by
Eq.(\ref{momentum}). In the non-relativistic limit, it has
$k_R^i=(2ME_{kin}-2MU^i_{sep})^{1/2}$ with $i=n,p$. The $k^{n}_R$ and
$k^{p}_R$ can have considerable difference due to the non-negligible
$U_{sym}$, which, for instance, can be found in
Refs.~\cite{liba04,fuchs05}, at various isospin asymmetries. In fact,
this is in also related to the Pauli blocking which contributes a
$\r^{2/3}$-dependence in the kinetic part of the symmetry
energy~\cite{ha86,mu87,chen01,be04,li06}. At low energies the
different Pauli blocking for neutrons and protons in asymmetric
matter are important. They can essentially lead to the dependence of
the in-medium NN cross sections on the isospin asymmetry of the
medium. In the energy region where RIA is valid, the kinetic energies
are much higher (roughly above 400MeV) with the relative nucleon
momentum $k\gg k_F$. The Pauli blocking thus becomes irrelevant.

\begin{figure}[tbh]
\begin{center}
\vspace*{-10mm}
\includegraphics[width=10cm,height=10cm]{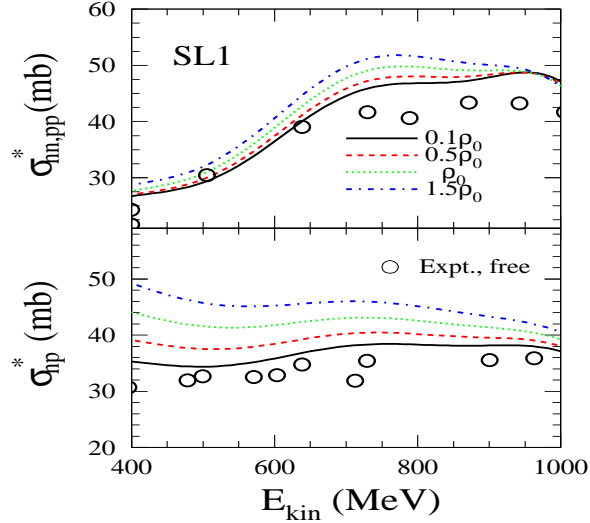}
\end{center}
\caption{(Color online) The in-medium NN cross sections as a function
of the  nucleon kinetic energy for different densities. The data for
free NN cross sections are taken from Ref.~\cite{li97}.}\label{f:f3}
\end{figure}

\begin{figure}[tbh]
\begin{center}
\vspace*{-10mm}
\includegraphics[width=10cm,height=10cm]{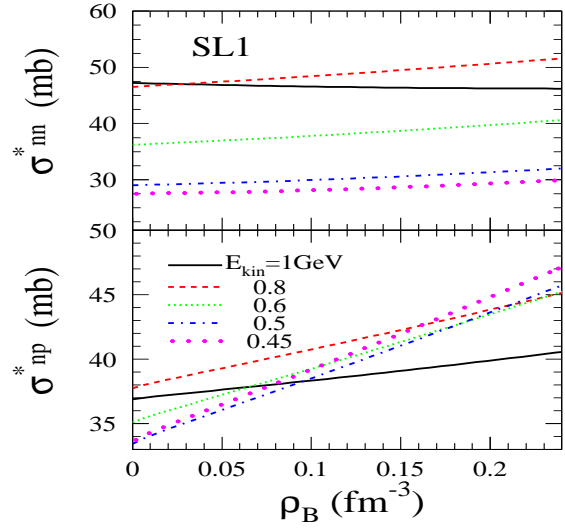}
\end{center}
\caption{(Color online) The in-medium NN cross sections as a function
of the density for various kinetic energies.}\label{f:f4}
\end{figure}
The $\sg^*_{np}$ and $\sg^*_{nn,pp}$ are calculated from the nucleon
MFP's with Eqs.(\ref{mfp3}) and (\ref{imcs}). As shown in
Figs.~\ref{f:f3} and \ref{f:f4}, the in-medium NN cross sections
increase with the density. At first glance, this feature seems to be
in contrast to that occurring at lower energies where the Pauli
blocking plays an important role in reducing the in-medium NN cross
sections. However, at higher kinetic energies the in-medium NN cross
sections are not necessarily constrained to be a descending function
of density. In Fig.~\ref{f:f3}, as a check we see that the calculated
NN cross sections at the very low density of $0.1\r_0$ are very close
to the free-space NN scattering data. Moreover, the in-medium NN
cross sections are shown to be linear in density. This can be easily
understood. As seen in Eqs.(\ref{mfp4}) and (\ref{imcs}), the
in-medium NN cross section are eventually determined by the SEP's
which are quadratic in relativistic optical potentials, while the
latter, given by Eq.(\ref{optr}), is linear in density. We note that
the scalar density in Eq.(\ref{optr}) is linear in density in the
relatively low density region where the RIA is applicable.
Considering the density factor appearing in the denominator of
Eq.(\ref{imcs}), the in-medium NN cross section is therefore linearly
proportional to the density. Interestingly, the ascending tendency of
the in-medium cross section with density has also be seen at higher
energies using some other approaches~\cite{li93,fu01,zh07}. For
instance, in Ref.~\cite{zh07}, with the closed time path Green's
function approach it was found that the $\sg^*_{nn,pp}$ increases
with density at $E_{kin}\ge 240$ MeV.

\begin{figure}[tbh]
\begin{center}
\vspace*{-10mm}
\includegraphics[width=10cm,height=10cm]{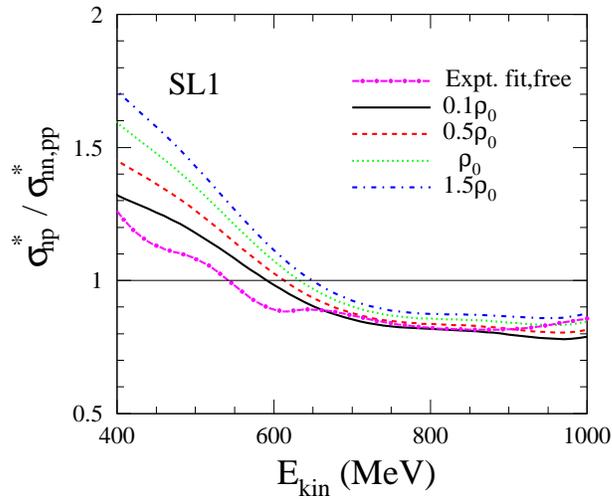}
\end{center}
\caption{ (Color online) The ratio of $\sg_{np}^*$ to $\sg_{nn,np}^*$ for various
densities. }\label{f:f5}
\end{figure}
It is also seen in Figs.~\ref{f:f3} and ~\ref{f:f4} that
$\sg_{np}^*$ depends more strongly on the density than $\sg_{nn,
pp}^*$, while the latter is more sensitive to the kinetic energy
than the former. This different density and energy dependence for
$\sg_{np}^*$ and $\sg_{nn,pp}^*$ is actually a result of the
distinction between the isovector (T=1) and isoscalar (T=0)
interactions. Shown in Fig.~\ref{f:f5} is the ratio
$\sg_{np}^*/\sg_{nn,pp}^*$ as a function of nucleon kinetic
energy. The ratio is an ascending function of density. This
feature is also different from that at low and intermediate
energies. It is clearly shown that the magnitude of
$\sg_{nn,pp}^*$ exceeds $\sg_{np}^*$ around $E_{kin}\ge 600$MeV
depending on the density. For a comparison, the experimental value
in free space is also shown in Fig. 6. The experimental
$\sg_{np}^*/\sg_{nn,pp}^*$ ratio becomes smaller than 1 above
about 580 MeV. The general trend is qualitatively consistent with
the calculations in the medium at the low density limit. It is
worth noting that the in-medium NN cross sections obtained in the
RIA rely little on the phenomenological freedoms except for those
in the RMF model to calculate the scalar density. In fact, within
the density region we considered in this study the scalar
densities are almost the same for the SL1 model and the best-fit
models, such as, the FSUGold~\cite{fsu}.

\section{Summary}
\label{summary}

We have evaluated the relativistic optical potentials for neutrons
and protons in asymmetric nuclear matter at high energies based on
the relativistic impulse approximation with the empirical NN
scattering amplitudes and the nuclear densities obtained in the
relativistic mean field model. The Schr\"odinger equivalent
potentials obtained from the relativistic optical potentials are
then applied to calculate the nucleon mean free paths and the
in-medium NN cross sections. The isospin-splitting of the proton
and neutron mean free paths in asymmetric nuclear matter changes
its sign at certain high energies, consistent with the changing
sign of the underlying imaginary part of the symmetry potential.
The small imaginary part of the symmetry potential at higher
energies, compared to the counterpart of the isoscalar
Schr\"odinger equivalent potential, accounts for the nucleon mean
free paths with a much reduced sensitivity to the isospin
asymmetry, while the negligibly small real part of the symmetry
potential results in the in-medium NN cross sections independent
of the isospin asymmetry of the medium. Moreover, in the energy
range where the RIA is valid the in-medium NN cross sections are
found to increase linearly with density.

\section*{Acknowledgement}

We thank G. C. Yong and P. Krastev for useful discussions. The work
was supported in part by the US National Science Foundation under
Grant No. PHY-0652548, the Research Corporation Award No. 7123, the National Natural
Science Foundation of China under Grant Nos. 10405031, 10575071 and
10675082, MOE of China under project NCET-05-0392, Shanghai
Rising-Star Program under Grant No. 06QA14024, the SRF for ROCS, SEM
of China,  the Knowledge Innovation Project of the Chinese Academy of
Sciences under Grant No. KJXC3-SYW-N2, and the China Major State
Basic Research Development Program under Contract No. 2007CB815004.

\end{document}